\documentclass[11pt]{article}
\usepackage{}
\usepackage{amsfonts}
\usepackage{amsmath,amssymb}
\usepackage{graphicx}
\usepackage{caption2}
\usepackage{epsfig}
\usepackage{subfigure}
\usepackage[dvips]{color}

\def\dref#1{(\ref{#1})}
\def\rm{\mathrm}
\begin{document}

\begin{center}
{\LARGE\bf{Global $H_\infty$ Consensus of Multi-Agent Systems with
Lipschitz Nonlinear Dynamics} \footnote[1] {Zhongkui Li, Xiangdong Liu and Mengyin Fu
are with the School of Automation, Beijing Institute of
Technology,\, Beijing 100081, P. R. China (e-mail:
zhongkli@gmail.com).
Lihua Xie is with the School of Electrical and
Electronic Engineering, Nanyang Technological University, 639798, Singapore
(e-mail: elhxie@ntu.edu.sg).}}
\end{center}
\vskip 0.3cm \centerline{Zhongkui Li, Xiangdong Liu, Mengyin Fu,
Lihua Xie} \vskip 1cm

{\noindent \small {\bf  Abstract}: This paper addresses the global
consensus problems of a class of nonlinear multi-agent systems with
Lipschitz nonlinearity and directed communication graphs,
by using a distributed consensus protocol
based on the relative states of neighboring agents. A two-step
algorithm is presented to construct a protocol, under which a
Lipschitz multi-agent system without disturbances can reach global
consensus for a strongly connected directed communication graph.
Another algorithm is
then given to design a protocol which can achieve global consensus
with a guaranteed $H_\infty$ performance for a Lipschitz multi-agent
system subject to external disturbances. The case with
a leader-follower communication graph is also discussed.
Finally, the effectiveness of the
theoretical results is demonstrated through a network of single-link
manipulators.

\vskip 0.2cm

}

\vskip 0.6cm
\section{Introduction}

In recent years, consensus control of multi-agent
systems has been an emerging topic in the systems
and control community and
has been extensively studied by numerous
researchers from various perspectives,
due to its potential applications in such broad areas as
satellite formation flying,
sensor networks, surveillance and reconnaissance
\cite{olfati-saber2007consensus,ren2007information}. 
A theoretical explanation is provided in
\cite{jadbabaie2003coordination} for the behavior observed in
the Vicsek Model \cite{vicsek1995novel} by using graph theory.
A general framework of the
consensus problem for networks of dynamic agents with fixed and
switching topologies is addressed in
\cite{olfati-saber2004consensus}. The conditions given by
\cite{jadbabaie2003coordination,olfati-saber2004consensus} are
further relaxed in \cite{ren2005consensus}. 
The controlled
agreement problem for multi-agent networks is considered from a
graph-theoretic perspective in \cite{rahmani2008controllability}.
Distributed $H_\infty$ consensus
and control problems are investigated in
\cite{lin2008distributed,li2011hinf} for networks of agents subject to
external disturbances.
Sampled-data control protocols are proposed to
achieve consensus for fixed and switching agent networks in
\cite{cao2009sampled,gao2009consensus}. The consensus problem of
networks of second- and high-order integrator agents is studied in
\cite{ren2008consensus,lin2009further,ren2007high-order}.
Different static and dynamic consensus protocols are proposed in
\cite{li2010consensus,li2011dynamic,scardovi2009synchronization,seo2009consensus}
for multi-agent systems with general linear dynamics. In the aforementioned
works, the agent dynamics are restricted to be
linear, where in some cases are simple integrators.

Recently, the consensus problems in networks with nonlinear dynamics
have been studied in \cite{arcak2007passivity,das2010distributed,yu2010second,song2010second,li2011global}.
Specifically, a passivity-based
design framework is proposed in \cite{arcak2007passivity}
to deal with the group coordination problem
with undirected communication graphs.
\cite{li2011global} studies
the global leader-follower consensus of
coupled Lur'e systems with certain sector-bound nonlinearity,
where the subgraph associated with the followers is
required to be undirected.
Neural adaptive tracking control of first-order nonlinear systems
with unknown dynamics is investigated in \cite{das2010distributed}.
The global consensus problems with and without a leader
are addressed in
\cite{yu2010second,song2010second}
for second-order multi-agent
systems with Lipschitz nonlinearity. 

In this paper, we extend to consider the global consensus problems of
high-order multi-agent systems with Lipschitz nonlinearity and
directed communication graphs. A distributed consensus
protocol is proposed, based on the relative states of neighboring
agents. A two-step algorithm is
presented to construct a protocol, under which a Lipschitz
multi-agent system without disturbances can reach global consensus for
a strongly connected directed communication graph.
For the case where the agents are subject to external disturbances, another algorithm is
then given to design a protocol which achieves global consensus
with a guaranteed $H_\infty$ performance for a strongly connected balanced communication graph.
It is worth mentioning that in these two algorithms the feedback gain design
of the consensus protocol is decoupled from the communication graph.
Finally, we further extend the results to the case with a leader-follower
communication graph which contains a directed spanning tree
with the leader as the root. Compared
to \cite{li2011global}, the requirement for the communication graph
is much relaxed in this paper.
Contrary to \cite{yu2010second,song2010second}
where the agent dynamics are restricted to be second-order,
the results derived in the current paper are applicable to any high-order Lipschitz
nonlinear multi-agent system and the global $H_\infty$
consensus problem is also studied here.


The rest of this paper is organized as follows. Some basic notation
and useful results of the graph theory are reviewed in Section 2.
The global consensus problem of multi-agent systems without
disturbances is considered in Section 3. The global $H_\infty$
consensus problem for agents subject to external disturbances is
investigated in Section 4. Extensions to the case with a
leader-follower graph are discussed in Section 5. A network of
single-link manipulators is utilized in
Section 6 to illustrate the analytical results.
Conclusions are drawn in Section 7.

\section{Concepts and Notation}

Let $\mathbf{R}^{n\times n}$ be the set of $n\times n$ real
matrices. The superscript $T$ means the transpose for real matrices.
$I_N$ represents the identity matrix of dimension $N$. Matrices, if
not explicitly stated, are assumed to have compatible dimensions.
Denote by $\mathbf{1}$ a column vector with all entries equal to
one. $\|\cdot\|$ refers to the Euclidean norm for vectors.
$\rm{diag}(A_1,\cdots,A_n)$ represents a block-diagonal matrix with
matrices $A_i,i=1,\cdots,n,$ on its diagonal. The matrix inequality
$A>B$ (respectively, $A\geq B$) means that $A-B$ is positive
definite (respectively, positive semidefinite). $A\otimes B$ denotes
the Kronecker product
of matrices $A$ and $B$. 
Denote by $\mathcal {L}_2[0,\infty)$ the space of square integrable
functions over $[0,\infty)$.

A directed graph $\mathcal {G}$ is a pair $(\mathcal {V}, \mathcal
{E})$, where $\mathcal {V}=\{v_1,\cdots,v_N\}$ is a nonempty finite
set of nodes and $\mathcal {E}\subseteq\mathcal {V}\times\mathcal
{V}$ is a set of edges, in which an edge is represented by an
ordered pair of distinct nodes. For an edge $(v_i,v_j)$, node $v_i$
is called the parent node, node $v_j$ the child node, and $v_i$ is a
neighbor of $v_j$. A graph with the property that
$(v_i,v_j)\in\mathcal {E}$ implies $(v_j, v_i)\in\mathcal {E}$ is
said to be undirected. A directed path from node $v_{i_1}$ to node $v_{i_l}$
is a sequence of ordered edges of the form $(v_{i_k}, v_{i_{k+1}})$,
$k=1,\cdots,l-1$. 
A directed
graph contains a directed spanning tree if there exists a node
called the root, which has no parent node, such that the node
has a directed path to every other node in the graph.
A directed graph is strongly connected if there is a directed
path from every node to every other node.
A directed graph has a directed spanning tree if it is strongly connected,
but not vice versa.

The adjacency matrix $\mathcal {A}=[a_{ij}]\in\mathbf{R}^{N\times
N}$ associated with the directed graph $\mathcal {G}$ is defined by
$a_{ii}=0$, $a_{ij}=1$ if $(v_j,v_i)\in\mathcal {E}$ and $a_{ij}=0$
otherwise. A directed graph is balanced if $\sum_{j=1}^Na_{ij}=\sum_{j=1}^Na_{ij}$
for all $i$.
The Laplacian matrix $\mathcal {L}=[\mathcal
{L}_{ij}]\in\mathbf{R}^{N\times N}$ is defined as $\mathcal
{L}_{ii}=\sum_{j\neq i}a_{ij}$ and $\mathcal {L}_{ij}=-a_{ij}$,
$i\neq j$. For an undirected graph, both $\mathcal {A}$ and $\mathcal {L}$
are symmetric.

{\bf Lemma 1}
\cite{olfati-saber2004consensus,ren2005consensus}. Zero is an
eigenvalue of $\mathcal {L}$ with $\mathbf{1}$ as a
right eigenvector and all other eigenvalues have positive real parts.
Furthermore, zero is a simple eigenvalue of $\mathcal {L}$ if and only if
the graph $\mathcal {G}$ has a directed spanning tree.

{\bf Lemma 2} \cite{qu2009cooperative,yu2010second}.
Suppose that $\mathcal {G}$ is strongly connected.
Let $r=[r_1,\cdots,r_N]^T$ be the positive left eigenvector of $\mathcal
{L}$ associated with the zero eigenvalue. Then, $R\mathcal {L}+\mathcal {L}^TR\geq 0$,
where $R={\rm{diag}}(r_1,\cdots,r_N)$.

{\bf Lemma 3} \cite{yu2010second}.
For a strongly connected graph $\mathcal {G}$ with Laplacian matrix $\mathcal {L}$,
define
its generalized algebraic connectivity as
$a(\mathcal {L})=\underset{r^Tx=0,x\neq0}{\min}\frac{x^T(R\mathcal {L}+\mathcal {L}^TR)x}{2x^TRx}$,
where $r$ and $R$ are defined as in Lemma 2. Then, $a(\mathcal {L})>0$.
For balanced graphs, $a(\mathcal {L})=\lambda_2(\frac{\mathcal {L}+\mathcal {L}^T}{2})$,
where $\lambda_2(\frac{\mathcal {L}+\mathcal {L}^T}{2})$ denotes
the smallest nonzero eigenvalue of $\frac{\mathcal {L}+\mathcal {L}^T}{2}$.


\section{Global Consensus without Disturbances}

Consider a group of $N$ identical nonlinear agents, described by
\begin{equation}\label{lip}
\begin{aligned}
\dot{x}_i =Ax_i+D_1f(x_i)+Bu_i, \quad i=1,\cdots,N,
\end{aligned}
\end{equation}
where $x_i\in\mathbf{R}^n$, $u_i\in\mathbf{R}^{p}$ are the state and
the control input of the $i$-th agent, respectively, $A$, $B$,
$D_1$, are constant matrices with compatible dimensions, and the
nonlinear function $f(x_i)$ is assumed to satisfy the Lipschitz
condition with a Lipschitz constant $\alpha>0$, i.e.,
\begin{equation}\label{lipcon}
\|f(x)-f(y)\|\leq \alpha\|x-y\|,\quad \forall\,x,y\in\mathbf{R}^n.
\end{equation}

The communication graph among the $N$ agents is represented by a
directed graph $\mathcal {G}$. It is supposed that each agent has access to the relative states
with respect to its neighbors. In order to achieve consensus, the
following distributed consensus protocol is proposed:
\begin{equation}\label{cl}
u_i=cK\sum_{j=1}^Na_{ij}(x_i-x_j), \quad i=1,\cdots,N,
\end{equation}
where $c>0\in\mathbf{R}$ denotes the coupling strength, $K\in\mathbf{R}^{p\times
n}$ is the feedback gain matrix, and $\mathcal
{A}=[a_{ij}]_{N\times N}$ is the adjacency matrix associated with
$\mathcal {G}$.

The objective is to design a consensus protocol \dref{cl} such that
the $N$ agents in \dref{lip} can achieve global consensus in the
sense of $\lim_{t\rightarrow \infty}\|x_i(t)- x_j(t)\|=0,$ $
\forall\,i,j=1,\cdots,N.$

%

Let $r=[r_1,\cdots,r_N]^T$ be the left eigenvector
of $\mathcal {L}$ associated with the zero eigenvalue, satisfying $r^T{\bf 1}=1$ and $r_i>0$, $i=1,\cdots,N$.
Define $e_i=x_i-\sum_{j=1}^N r_jx_j$ and $e=[e_1^T,\cdots,e_N^T]^T$. Then, we
get
\begin{equation}\label{conerr}
\begin{aligned}
e=[(I_{N}-{\bf 1}r^T)\otimes I_n]x,
\end{aligned}
\end{equation}
where
$x=[x_1,\cdots,x_N]^T$. Clearly, $e$ satisfies $(r^T\otimes I_n)e=0$,
i.e., $\sum_{j=1}^N r_je_j=0$.

By the definition of $r$, it is easy
to see that $0$ is a simple eigenvalue of
$I_N-\mathbf{1}r^T$ with $\mathbf{1}$ as a
right eigenvector, and 1 is the other eigenvalue with
multiplicity $N-1$. Then, it follows from \dref{conerr} that $e=0$ if and only if
$x_1=\cdots=x_N$. Therefore, the consensus problem under the protocol
\dref{cl} can be reduced to the
asymptotical stability of $e$. Using \dref{cl} for \dref{lip}, it
can be verified that $e$ satisfies the following
dynamics:
\begin{equation}\label{nete}
\begin{aligned}
\dot{e}_i
=Ae_i+D_1f(x_i)-D_1\sum_{j=1}^{N}r_if(x_j)+c\sum_{j=1}^{N}\mathcal
{L}_{ij}BKe_j, \quad i=1,\cdots,N,
\end{aligned}
\end{equation}
where $\mathcal {L}=[\mathcal {L}_{ij}]_{N\times N}$ is the
Laplacian matrix associated with $\mathcal {G}$.

Next, an algorithm is presented to select the control parameters in
\dref{cl}.

{\bf Algorithm 1}. For the agents in \dref{lip}, a consensus
protocol \dref{cl} can be constructed as follows:
\begin{itemize}
\item[1)]
Solve the following LMI:
\begin{equation}\label{t41}
\begin{bmatrix} AP+PA^T- \tau BB^T+\alpha^2 D_1 D_1^T & P
    \\P & -I\end{bmatrix}<0,
\end{equation}
to get a matrix $P>0$ and a scalar $\tau>0$.
Then, choose $K=-\frac{1}{2}B^TP^{-1}$.

\item[2)] Select the coupling strength
$c\geq\frac{\tau}{a(\mathcal {L})}$, where
$a(\mathcal {L})$ is the generalized algebraic connectivity
of $\mathcal {G}$, defined as in Lemma 3.
\end{itemize}


The following presents a sufficient condition for the global
consensus of \dref{nete}.

{\bf Theorem 1}. Assume that the directed
graph $\mathcal {G}$ is strongly connected and
there exists a solution to \dref{t41}. Then, the $N$ agents in
\dref{lip} can reach global consensus under the protocol \dref{cl}
constructed by Algorithm 1.

{\bf Proof.} Consider the Lyapunov function candidate
$$V_1=\sum_{i=1}^{N}r_ie_i^TP^{-1}e_i,$$
where $r=[r_1,\cdots,r_N]^T$ is defined as in \dref{conerr}.
Clearly $V_1$ is positive definite.
The time derivative of $V_1$ along the trajectory of \dref{nete} is
given by
\begin{equation}\label{lyae}
\begin{aligned}
\dot{V}_1 &=2\sum_{i=1}^{N}r_ie_i^TP^{-1}[
Ae_i+D_1f(x_i)-D_1\sum_{j=1}^{N}r_if(x_j)+c\sum_{j=1}^{N}\mathcal
{L}_{ij}BKe_j]\\
&=2\sum_{i=1}^{N}r_ie_i^TP^{-1}[Ae_i+c\sum_{j=1}^{N}\mathcal
{L}_{ij}BKe_j]\\
&\quad +2\sum_{i=1}^{N}r_ie_i^TP^{-1}D_1[
f(x_i)-f(\bar{x})+f(\bar{x})-\sum_{j=1}^{N}r_if(x_j)],
\end{aligned}
\end{equation}
where $\bar{x}=\sum_{j=1}^N r_jx_j$.

Using the Lipschitz condition \dref{lipcon} gives
\begin{equation}\label{lipc2}
\begin{aligned}
2e_i^TP^{-1}D_1(f(x_i)-f(\bar{x}))&\leq
2\alpha\|e_i^TP^{-1}D_1\|\|e_i\| \\&\leq
e_i^T(\alpha^2P^{-1}D_1 D_1^TP^{-1}+I)e_i.
\end{aligned}
\end{equation}
Since
$\sum_{i=1}r_ie_i=0$, we have
\begin{equation}\label{lipc3}
\sum_{i=1}r_ie_i^TP^{-1}D_1[f(\bar{x})-\frac{1}{N}\sum_{j=1}^{N}f(x_j)]=0.
\end{equation}

Let $\xi_i=P^{-1}e_i$,
$\xi=[\xi_1^T,\cdots,\xi_N^T]^T$. Substituting $K=-B^TP^{-1}$ in \dref{lyae},
it follows from \dref{lyae} by using \dref{lipc2} and \dref{lipc3} that
\begin{equation}\label{lyae2}
\begin{aligned}
\dot{V}_1 &\leq
\sum_{i=1}^{N}r_i\xi_i^T[(AP+PA^T+\alpha^2D_1 D_1^T+P^2)\xi_i+2c\sum_{j=1}^{N}\mathcal
{L}_{ij}BKP\xi_j]\\
&=\xi^T[R\otimes (AP+PA^T+\alpha^2D_1
D_1^T+P^2)-\frac{c}{2}(R\mathcal {L}+\mathcal {L}^TR)\otimes (BB^T)]\xi,
\end{aligned}
\end{equation}
where $R={\rm{diag}}(r_1,\cdots,r_N)$.

Since $(r^T\otimes I_n)e=0$,
i.e., $(r^T\otimes I_n)\xi=0$, we can get from Lemma 3 that
\begin{equation}\label{rich}
\xi^T[(R\mathcal {L}+\mathcal {L}^TR)\otimes I_n]\xi\geq 2a(\mathcal {L})\xi^T(R\otimes I_n)\xi,
\end{equation}
where $a(\mathcal {L})>0$. In light of \dref{rich}, it then follows from \dref{lyae2}
that
\begin{equation}\label{lyae3}
\begin{aligned}
\dot{V}_1
&\leq \xi^T[I_N\otimes (AP+PA^T+\alpha^2D_1 D_1^T+P^2-c\,a(\mathcal {L})BB^T)]\xi.
\end{aligned}
\end{equation}
By steps 1) and 2) in Algorithm 1, we can obtain that
\begin{equation}\label{lmi1}
\begin{aligned}
&AP+PA^T+\alpha^2D_1 D_1^T+P^2-c\,a(\mathcal {L})BB^T\\
&\qquad \qquad\leq AP+PA^T+\alpha^2D_1 D_1^T+P^2-\tau BB^T<0,
\end{aligned}
\end{equation}
where the last inequality follows from \dref{t41} by using the
Schur complement lemma \cite{boyd1994linear}. Therefore, it follows
from \dref{lyae3} that $\dot{V}_1<0$, implying
that $e(t)\rightarrow 0$, as
$t\rightarrow \infty$. That is,
the global consensus of network \dref{nete} is achieved. \hfill
$\blacksquare$

{\bf Remark 1}. Theorem 1 converts the global consensus of
the $N$ Lipschitz agents in \dref{lip} under the protocol \dref{cl}
to the feasibility of a low-dimensional linear matrix inequality.
The effects of the communication topology on the stability of
consensus are characterized by the generalized algebraic connectivity
of
the corresponding Laplacian matrix $\mathcal {L}$. 
It is worth mentioning that global consensus problems of
multi-agent systems with Lipschitz nonlinearity were also considered in
\cite{yu2010second,song2010second}. However, the agent dynamics are
restricted to be second-order there. By contrast, The results
given in this section are applicable to any high-order
Lipschitz nonlinear multi-agent system. 

{\bf Remark 2}. By using Finsler's Lemma
\cite{iwasaki1994all}, it is not difficult to get that there exist
a $P>0$ and a $\tau>0$ such that \dref{t41} holds if
and only if there exists a $K$ such that
$(A+BK)P+P(A+BK)^T+\alpha^2D_1D_1^T+P^2<0$, which
with $D_1=I$ is dual to the observer design problem for a single
Lipschitz system in \cite{rajamani1998observers}. Hence, the
sufficient condition for the existence of the observer for a
Lipschitz system in
\cite{rajamani1998observers,rajamani1998existence}
can be used to check the solvability of the LMI \dref{t41}. 

{\bf Remark 3}. It is worth
noting that Algorithm 1 has a favorable decoupling feature.
Specifically, the first step deals only with the
agent dynamics in \dref{lip}
while the second step tackles the communication graph
by adjusting the coupling strength $c$.
The generalized algebraic connectivity
$a(\mathcal {L})$ for
a given graph can be obtained by using Lemma 8 in \cite{yu2010second}.
For a large-scale graph, for which $a(\mathcal {L})$ might be
not easy to compute, we can
simply choose the coupling strength to be large enough.
According to Lemma 3,
$a(\mathcal {L})$ can be replaced by
$\lambda_2(\frac{\mathcal {L}+\mathcal {L}^T}{2})$
in step 2), if
the directed graph $\mathcal {L}$ is balanced and strongly connected.

\section{Global $H_\infty$ Consensus with Disturbances}

This section continues to consider a network of $N$ identical nonlinear
agents subject to external disturbances, described by
\begin{equation}\label{d1}
\begin{aligned}
    \dot{x}_i &=Ax_i+D_1f(x_i)+Bu_i+D_2\omega_i,\quad i=1,\cdots,N,
\end{aligned}
\end{equation}
where $x_i\in\mathbf{R}^n$ is the state of the $i$-th agent,
$u_i\in\mathbf{R}^{p}$ is the control input, $\omega_i\in\mathcal
{L}_2^{m_1}[0,\infty)$ is the external disturbance, and $f(x_i)$
is a nonlinear function satisfying \dref{lipcon}.
The communication graph $\mathcal {G}$ is assumed
to balanced and strongly connected in this section.

The objective in this section is to design a protocol \dref{cl} for
the agents in \dref{d1} to reach global consensus and meanwhile
maintain a desirable disturbance rejection performance. To this end,
define the performance variable $z_i$, $i=1,2,\cdots,N$, as the
average of the weighted relative states of the agents:
\begin{equation}\label{pv}
z_i=\frac{1}{N}\sum_{j=1}^N C(x_i-x_j), \quad i=1,\cdots,N,
\end{equation}
where $z_i\in\mathbf{R}^{m_2}$ and $C\in\mathbf{R}^{m_2\times n}$ is
a constant matrix.

With \dref{d1}, \dref{cl}, and \dref{pv}, the closed-loop network
dynamics can be obtained as
\begin{equation}\label{neted}
\begin{aligned}
\dot{x}_i
&=Ax_i+D_1f(x_i)+c\sum_{j=1}^{N}\mathcal
{L}_{ij}BKx_j+D_2\omega_i, \\
z_i &=\frac{1}{N}\sum_{j=1}^N C(x_i-x_j),\qquad i=1,\cdots,N,
\end{aligned}
\end{equation}

The global $H_\infty$ consensus problem for \dref{d1} under the
protocol \dref{cl} is first defined.

{\bf Definition 1}. Given a positive scalar $\gamma$, the
protocol \dref{cl} is said to achieve global consensus with a
guaranteed $H_\infty$ performance $\gamma$ for the agents in
\dref{d1}, if the following two requirements hold:
\begin{itemize}
\item[(1)] The network
\dref{neted} with $\omega_i=0$ can reach global consensus in the sense of
$\lim_{t\rightarrow \infty} \|x_i-x_j\|=0,$
$\forall\,i,j=1,\cdots,N$.

\item[(2)] Under the zero-initial condition, the performance variable $z$
satisfies
\begin{equation}\label{perf}
J=\int_0^\infty [z^T(t)z(t)-\gamma^2
\omega^T(t)\omega(t)]dt<0,
\end{equation}
where $z=[z_1^T,\cdots,z_N^T]^T,$
$\omega=[\omega_1^T,\cdots,\omega_N^T]^T$.
\end{itemize}

Next, an algorithm for the protocol \dref{cl} is presented.

{\bf Algorithm 2}. For a given scalar $\gamma>0$ and the
agents in \dref{d1}, a consensus protocol \dref{cl} can be
constructed as follows:
\begin{itemize}
\item[1)]
Solve the following LMI:
\begin{equation}\label{t51}
\begin{bmatrix} AQ+QA^T- \epsilon BB^T+\alpha^2D_1 D_1^T & Q & QC^T & D_2\\
Q & -I& 0 &0 \\CQ &0 &-I & 0\\
D_2^T & 0 & 0 & -\gamma^2 I\end{bmatrix}<0
\end{equation}
to get a matrix $Q>0$ and a scalar $\epsilon>0$. Then, choose $K=-\frac{1}{2}B^TQ^{-1}$.

\item[2)] Select the coupling strength
$c\geq\frac{\epsilon}{\lambda_2(\frac{\mathcal {L}+\mathcal {L}^T}{2})}$,
where $\lambda_2(\frac{\mathcal {L}+\mathcal {L}^T}{2})$ denotes the smallest
nonzero eigenvalue of $\frac{\mathcal {L}+\mathcal {L}^T}{2}$.
\end{itemize}

{\bf Theorem 2}. Assume that $\mathcal {G}$ is balanced and strongly connected, and
there exists a solution to \dref{t51}. Then, the protocol \dref{cl}
constructed by Algorithm 2 can achieve global consensus with a
guaranteed $H_\infty$ performance $\gamma>0$ for the $N$ agents in
\dref{d1}.

{\bf Proof.} Let $e_i=x_i-\frac{1}{N}\sum_{j=1}^{N} x_j$, and
$e=[e_1^T,\cdots,e_N^T]^T$. As shown in the last section, we know
that $e=0$ if and only if $x_1=\cdots=x_N$. From \dref{neted}, it is
easy to get that $e$ satisfies the following dynamics:
\begin{equation}\label{neted2}
\begin{aligned}
\dot{e}_i &=
Ae_i+D_1f(x_i)-\frac{1}{N}\sum_{j=1}^{N}D_1f(x_j)\\
&\quad+c\sum_{j=1}^{N}\mathcal
{L}_{ij}BKe_j+\frac{1}{N}\sum_{j=1}^ND_2(\omega_i-\omega_j),\\
z_i &=Ce_i,\quad i=1,\cdots,N.
\end{aligned}
\end{equation}
Therefore, the protocol \dref{cl} solves the global $H_\infty$
consensus problem, if the system \dref{neted2} is asymptotically
stable and satisfies \dref{perf}.

Consider the Lyapunov function candidate
$$V_2=\sum_{i=1}^{N}e_i^TQ^{-1}e_i.$$
By following similar steps to those in Theorem 1, we can obtain the
time derivative of $V_2$ along the trajectory of \dref{neted2} as
\begin{equation}\label{lyaet22}
\begin{aligned}
\dot{V}_2
&\leq\sum_{i=1}^{N}e_i^T[(2Q^{-1}A+I+\alpha^2Q^{-1}D_1 D_1^TQ^{-1})e_i+2c
\sum_{j=1}^{N}\mathcal
{L}_{ij}Q^{-1}BKe_j\\
&\quad+\frac{2}{N}\sum_{j=1}^NQ^{-1}D_2(\omega_i-\omega_j)]\\
&=\hat{e}^T[I_N\otimes (AQ+QA^T+Q^2+\alpha^2D_1D_1^T)-\frac{c}{2}(\mathcal
{L}+\mathcal
{L}^T)\otimes
(BB^T)]\hat{e}\\
&\quad+2\hat{e}^T(\mathcal {M}\otimes D_2)\omega,
\end{aligned}
\end{equation}
where $\hat{e}_i=Q^{-1}e_i$,
$\hat{e}=[\hat{e}_1^T,\cdots,\hat{e}_N^T]^T$, and $\mathcal
{M}=I_N-\frac{1}{N}\mathbf{1}\mathbf{1}^T.$

Next, for any nonzero $\omega$, we have
\begin{equation}\label{lyaet23}
\begin{aligned}
J
&=\int_0^\infty [z^T(t)z(t)-\gamma^2 \omega^T(t)\omega(t)+\dot{V}_2]dt-V_2(\infty)+V_2(0)\\
&\leq\int_0^\infty \hat{e}^T[I_N\otimes
(AQ+QA^T+Q^2+\alpha^2D_1D_1^T)-\frac{c}{2}(\mathcal
{L}+\mathcal {L}^T)\otimes (BB^T)  \\
    &\qquad\qquad +I_N\otimes (QC^TCQ)+\gamma^{-2}\mathcal {M}^2\otimes (D_2D_2^T)]\hat{e}\,dt\\
    &\quad -\int_0^\infty
    \gamma^2[\omega -\gamma^{-2}(\mathcal {M}\otimes D_2^T)\hat{e}]^T
    [\omega -\gamma^{-2}(\mathcal {M}\otimes D_2^T)\hat{e}]dt\\
    &\quad-V_2(\infty)+V_2(0).
\end{aligned}
\end{equation}

By noting that ${\bf 1}$ and ${\bf 1}^T$ are the right and left
eigenvectors of $\mathcal {L}+\mathcal {L}^T$ associated with the zero eigenvalue,
respectively, we have
$$\mathcal {M}(\mathcal {L}+\mathcal {L}^T)=\mathcal {L}+\mathcal {L}^T=(\mathcal {L}+\mathcal {L}^T)\mathcal {M}.$$
Thus there exists a unitary matrix
$\hat{U}\in\mathbf{R}^{N\times N}$ such that $\hat{U}^T\mathcal
{M}\hat{U}$ and $\hat{U}^T(\mathcal {L}+\mathcal {L}^T)\hat{U}$ are both diagonal
\cite{horn1990matrix}. Since $\mathcal {L}+\mathcal {L}^T$ and $\mathcal {M}$ have
the same right and left eigenvectors corresponding to the zero
eigenvalue, namely, ${\bf 1}$ and ${\bf 1}^T$, we can choose
$\hat{U}=\left[\begin{smallmatrix} \frac{\mathbf{1}}{\sqrt{N}} & Y
\end{smallmatrix}\right],$
$\hat{U}^{T}=\left[\begin{smallmatrix}\frac{\mathbf{1}^T}{\sqrt{N}}\\
W \end{smallmatrix}\right]$, with $Y\in\mathbf{R}^{N\times(N-1)}$,
$W\in\mathbf{R}^{(N-1)\times N}$, satisfying
\begin{equation}\label{sd}
\begin{aligned}
\hat{U}^T\mathcal {M}\hat{U} &=\Pi=\begin{bmatrix} 0 & 0 \\ 0 & I_{N-1}\end{bmatrix},\\
\frac{1}{2}\hat{U}^T(\mathcal {L}+\mathcal {L}^T)\hat{U}
&=\Lambda=\rm{diag}(0,\lambda_2,\cdots,\lambda_N),
\end{aligned}
\end{equation}
where $\lambda_i$, $i=2,\cdots,N$, are the nonzero eigenvalues of $\frac{\mathcal {L}+\mathcal {L}^T}{2}$.
Let $\zeta\triangleq[\zeta_1^T,\cdots,\zeta_N^T]^T=(\hat{U}^T\otimes
I_n)\hat{e}$. Clearly,
$\zeta_1=(\frac{\mathbf{1}^T}{\sqrt{N}}\otimes Q^{-1})e=0$. By using
\dref{sd}, we can obtain that
\begin{equation}\label{lyaet24}
\begin{aligned}
&\hat{e}^T[I_N\otimes
(AQ+QA^T+Q^2+\alpha^2D_1D_1^T)-\frac{c}{2}(\mathcal {L}+\mathcal {L}^T)\otimes (BB^T)    \\
    &\quad+I_N\otimes (QC^TCQ)+\gamma^{-2}\mathcal {M}^2\otimes (D_2D_2^T)]\hat{e}\\
&\qquad=\zeta^T[I_N\otimes
(AQ+QA^T+Q^2+\alpha^2D_1D_1^T)-c\Lambda\otimes (BB^T)\\
&\qquad\quad+I_N\otimes (QC^TCQ)+\gamma^{-2}\Pi^2\otimes (D_2D_2^T)]\zeta\\
&\qquad=\sum_{i=2}^{N}\zeta_i^T[AQ+QA^T+Q^2+\alpha^2D_1D_1^T-c\lambda_iBB^T\\
&\qquad\quad+\gamma^{-2}D_2D_2^T+QC^TCQ]\zeta_i.
\end{aligned}
\end{equation}
In light of steps 1) and 2) in Algorithm 2, we have
\begin{equation}\label{lyaet25}
\begin{aligned}
&AQ+QA^T+Q^2+\alpha^2D_1D_1^T-c\lambda_iBB^T+\gamma^{-2}D_2D_2^T+QC^TCQ\\
&\qquad\leq AQ+QA^T+Q^2+\alpha^2D_1D_1^T-\epsilon
BB^T+\gamma^{-2}D_2D_2^T+QC^TCQ<0.
\end{aligned}
\end{equation}

By comparing Algorithm 2 with Algorithm 1, it follows from Theorem 1 that the
first condition in Definition 1 holds. Since $x(0)=0$, it is clear that $V_2(0)=0$.
Considering
\dref{lyaet24} and \dref{lyaet25}, we can obtain from \dref{lyaet23}
that $J<0$. Therefore, the global $H_\infty$ consensus problem is
solved. \hfill $\blacksquare$

{\bf Remark 4}. Theorem 2 and Algorithm 2 extend
Theorem 1 and Algorithm 1 to evaluate the performance of a
multi-agent network subject to external disturbances.
The decoupling property of Algorithm 1 as stated in Remark 3 still
holds for Algorithm 2.

\section{Extensions}

In the above sections, the communication graph is assumed to be
strongly connected, where the final consensus value reached by the agents
is generally not explicitly known. In many practical cases, it is
desirable that the agents' states asymptotically approach a
reference state. In this section, we extend to consider the case
where a network of $N$ agents in \dref{lip} maintains a leader-follower
communication graph $\mathcal {G}$.
An agent is called a
leader if the agent has no neighbor, i.e.,
it does not receive any information.
An agent is called a follower
if the agent has at least one neighbor.
Without loss of generality, assume that the agent indexed by
1 is the leader and the rest $N-1$ agents are followers.
The following distributed consensus protocol is proposed
for each follower:
\begin{equation}\label{clf}
u_i=cK\sum_{j=1}^Na_{ij}(x_i-x_j),\quad
i=2,\cdots,N,
\end{equation}
where $c>0$, $K\in\mathbf{R}^{p\times n}$, $a_{ij}$
are the same as defined in \dref{cl}.

The objective in this section is to solve the leader-follower global consensus problem, i.e.,
to design a consensus protocol \dref{clf} under which the states of
the followers asymptotically approach the state of the leader in the sense of
$\lim_{t\rightarrow
\infty}\|x_i(t)- x_1(t)\|=0$, $ \forall\,i=2,\cdots,N$.

In the sequel, we make the following assumption.

{\bf Assumption 1}.
The communication graph $\mathcal {G}$ contains a directed spanning tree
with the leader as the root.

%
%

Because the leader has no neighbors, the Laplacian matrix $\mathcal
{L}$ associated with
$\mathcal {G}$ can be partitioned as
\begin{equation}\label{lapc}
\mathcal {L}=\begin{bmatrix} 0 & 0_{1\times (N-1)} \\
\mathcal {L}_2 & \mathcal {L}_1\end{bmatrix},
\end{equation}
where $\mathcal
{L}_2\in\mathbf{R}^{(N-1)\times 1}$ and $\mathcal {L}_1\in\mathbf{R}^{(N-1)\times (N-1)}$.
Under Assumption 1, it follows from Lemma 1 that all the eigenvalues of $\mathcal {L}_1$
have positive real parts.

{\bf Lemma 4} \cite{qu2009cooperative,das2010distributed}.
Define
\begin{equation}\label{notlf}
\begin{aligned}
q &=[q_2,\cdots,q_{N}]^T=\mathcal {L}_1^{-1}{\bf 1},\\
G &=\rm{diag}(1/q_2,\cdots,1/q_{N}),~H=\frac{1}{2}(G\mathcal {L}_1+\mathcal {L}_1^TG).
\end{aligned}
\end{equation}
Then, both $G$ and $H$ are positive definite.

{\bf Algorithm 3}. For the agents in \dref{lip} satisfying Assumption 1, a consensus
protocol \dref{clf} can be constructed as follows:
\begin{itemize}
\item[1)]
Solve the following LMI:
$$\begin{bmatrix} AS+SA^T- \kappa BB^T+\alpha^2 D_1 D_1^T & S
    \\S & -I\end{bmatrix}<0,$$
to get a matrix $S>0$ and a scalar $\kappa>0$.
Then, choose $K=-\frac{1}{2}B^TS^{-1}$.

\item[2)] Select the coupling strength
$c\geq\frac{\kappa}{\lambda_1(H)\underset{i=2,\cdots,N}{\min}q_i}$, with
$\lambda_1(H)$ being the the smallest
    eigenvalue of $H$, where $H$ and $q_i$ are defined in \dref{notlf}.
\end{itemize}

{\bf Remark 5}. For the case where the subgraph associated with
the followers is balanced and strongly connected, $\mathcal {L}_1+\mathcal {L}_1^T>0$ \cite{li2010consensus}.
Then, by letting $G=I$, step 2) can be simplified to $c\geq\frac{\tau}{\lambda_1(\frac{\mathcal {L}_1+\mathcal {L}_1^T}{2})}$.

{\bf Theorem 3}. Suppose that Assumption 1 holds
and there exists a solution to \dref{t41}. Then, the
consensus protocol \dref{clf} given by Algorithm 3 solves
the leader-follower global consensus problem for
the $N$ agents described by \dref{lip}.

{\bf Proof}. Denote the consensus errors by $\upsilon_i=x_i-x_1$, $i=2,\cdots,N$.
Then, we can obtain from \dref{lip} and \dref{clf} the closed-loop network dynamics as
\begin{equation}\label{netd3}
\dot{\upsilon}_i  =
A\upsilon_i+D_1[f(x_i)-f(x_1)]+cBK\sum_{j=1}^{N}a_{ij}(\upsilon_i-\upsilon_j), \quad i=2,\cdots,N,
\end{equation}
Clearly, the leader-follower global consensus
problem can be reduced to the asymptotical stability of
\dref{netd3}.

Consider the Lyapunov function candidate
$$V_3=\sum_{i=2}^{N}q_i\upsilon_i^TS^{-1}\upsilon_i,$$
where $q_i$ is defined as in \dref{notlf}. Clearly $V_3$ is positive definite.
Following similar steps in proving Theorem 1,
the time derivative of $V_3$ along the trajectory of \dref{netd3} is
obtained as
\begin{equation}\label{lyaef}
\begin{aligned}
\dot{V}_3 &=2\sum_{i=2}^{N}q_i\upsilon_i^TS^{-1}[
A\upsilon_i+D_1(f(x_i)-f(x_1))+cBK\sum_{j=1}^{N}a_{ij}(\upsilon_i-\upsilon_j)]\\
&\leq
\sum_{i=2}^{N}q_i\tilde{\upsilon}_i^T[(AS+SA^T+\alpha^2D_1 D_1^T+S^2)\tilde{\upsilon}_i
-cBB^T\sum_{j=1}^{N}a_{ij}(\tilde{\upsilon}_i-\tilde{\upsilon}_j)]\\
&=\tilde{\upsilon}^T[G\otimes (AS+SA^T+\alpha^2D_1 D_1^T+S^2)-cH\otimes (BB^T)]\tilde{\upsilon}\\
&=\tilde{\upsilon}^T[G\otimes (AS+SA^T+\alpha^2D_1 D_1^T+S^2-\tau BB^T)]\tilde{\upsilon}\\
&\quad+\tilde{\upsilon}^T[(\tau G-cH)\otimes (BB^T)]\tilde{\upsilon},
\end{aligned}
\end{equation}
where $\tilde{\upsilon}_i=S^{-1}\upsilon_i$,
$\tilde{\upsilon}=[\tilde{\upsilon}_2^T,\cdots,\tilde{\upsilon}_{N}^T]^T$,
$G$, $H$ are defined in \dref{notlf}, and we
have used \dref{lipc2} to get the first inequality.

Since $G$ and $H$ are positive definite, we can get that
$\tau G-cH<0$ if
$\tau G<c\lambda_1(H)I$, which is true in light of step 2) in Algorithm 3.
Thus,  $(\tau G-cH)\otimes (BB^T)\leq0$. Then, it follows from
\dref{lmi1} and \dref{lyaef} that $\dot{V}_3<0$, implying that
\dref{netd3} is asymptotically stable. This completes the proof.
\hfill $\blacksquare$

The global $H_\infty$ consensus for the agents in \dref{d1} with a
leader-follower communication graph can be discussed similarly,
thereby is omitted here for brevity.


\section{Simulation Examples}

In this section, a simulation example is provided to validate the
effectiveness of the theoretical results.

\begin{figure}[htbp]
\centering
\includegraphics[width=0.3\linewidth]{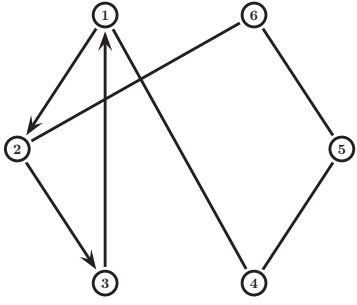}
\caption{The communication graph. }
\end{figure}

Consider a network of six single-link manipulators with revolute
joints actuated by a DC motor. The dynamics of the $i$-th
manipulator is described by \dref{d1}, with
\cite{rajamani1998existence,zhu2002note}
$$\begin{aligned}
x_{i} &=\begin{bmatrix}
x_{i1}\\x_{i2}\\x_{i3}\\x_{i4}\end{bmatrix},\quad
A=\begin{bmatrix}0 & 1& 0 &0\\ -48.6 & -1.26 & 48.6 & 0\\
0 &0 & 0 & 10\\ 1.95 & 0 & -1.95 &0\end{bmatrix},\quad B
=\begin{bmatrix} 0 \\ 21.6 \\ 0 \\ 0\end{bmatrix},
\\D_1&=I,\quad D_2=\begin{bmatrix} 0 & 1 & 0.4&
0\end{bmatrix}^T,\quad C=\begin{bmatrix} 1 & 0 & 0 &
0\end{bmatrix},\\f(x_i)&=\begin{bmatrix} 0 & 0& 0&
-0.333\mathrm{sin}(x_{i1})\end{bmatrix}^T.
\end{aligned}$$
Clearly, $f(x_i)$ here satisfies \dref{lipcon} with a Lipschitz
constant $\alpha=0.333$. The external disturbance here is
$\omega=[w, -w, 1.5w, 3w,-0.6w,2w]^T$, where $w(t)$ is a one-period square wave
starting at $t=0$ with width 2 and height 1.

\begin{figure}[htbp]\centering
\begin{minipage}[b]{0.4\linewidth} \centering
\includegraphics[width=\linewidth,height=0.5\linewidth]{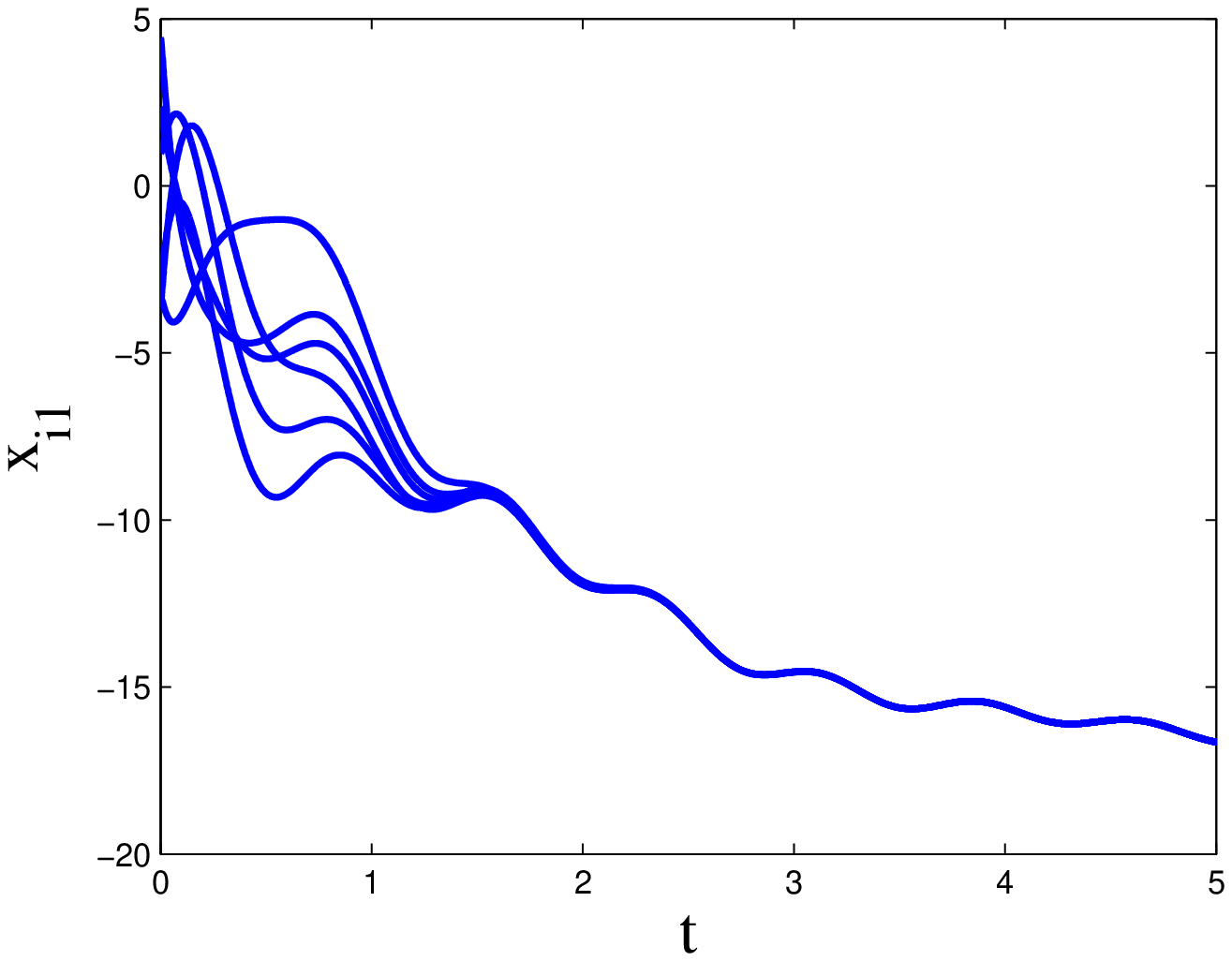}
\end{minipage}%
\begin{minipage}[b]{0.4\linewidth} \centering
\includegraphics[width=\linewidth,height=0.5\linewidth]{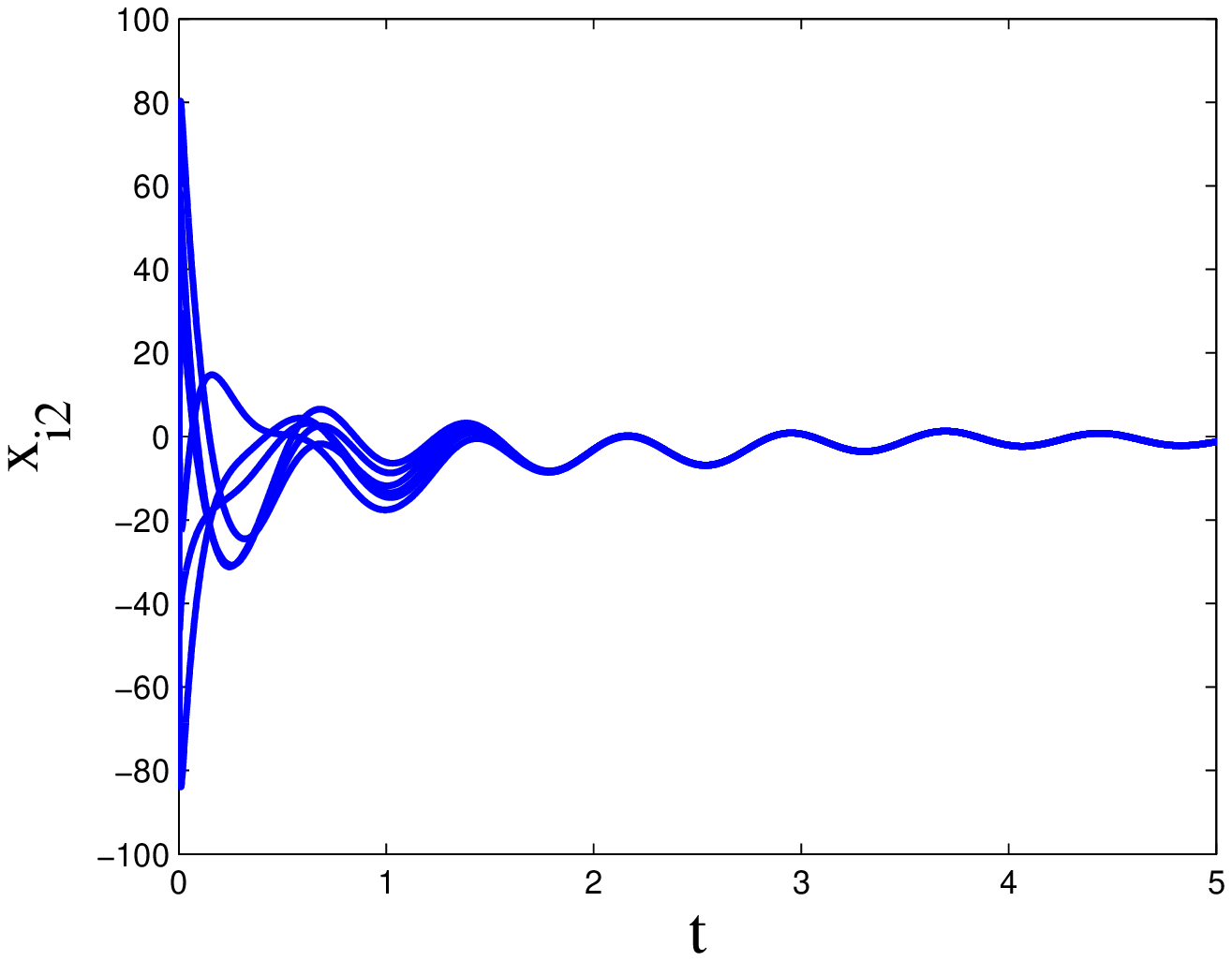}
\end{minipage}\\
 \begin{minipage}[b]{0.4\linewidth} \centering
\includegraphics[width=\linewidth,height=0.5\linewidth]{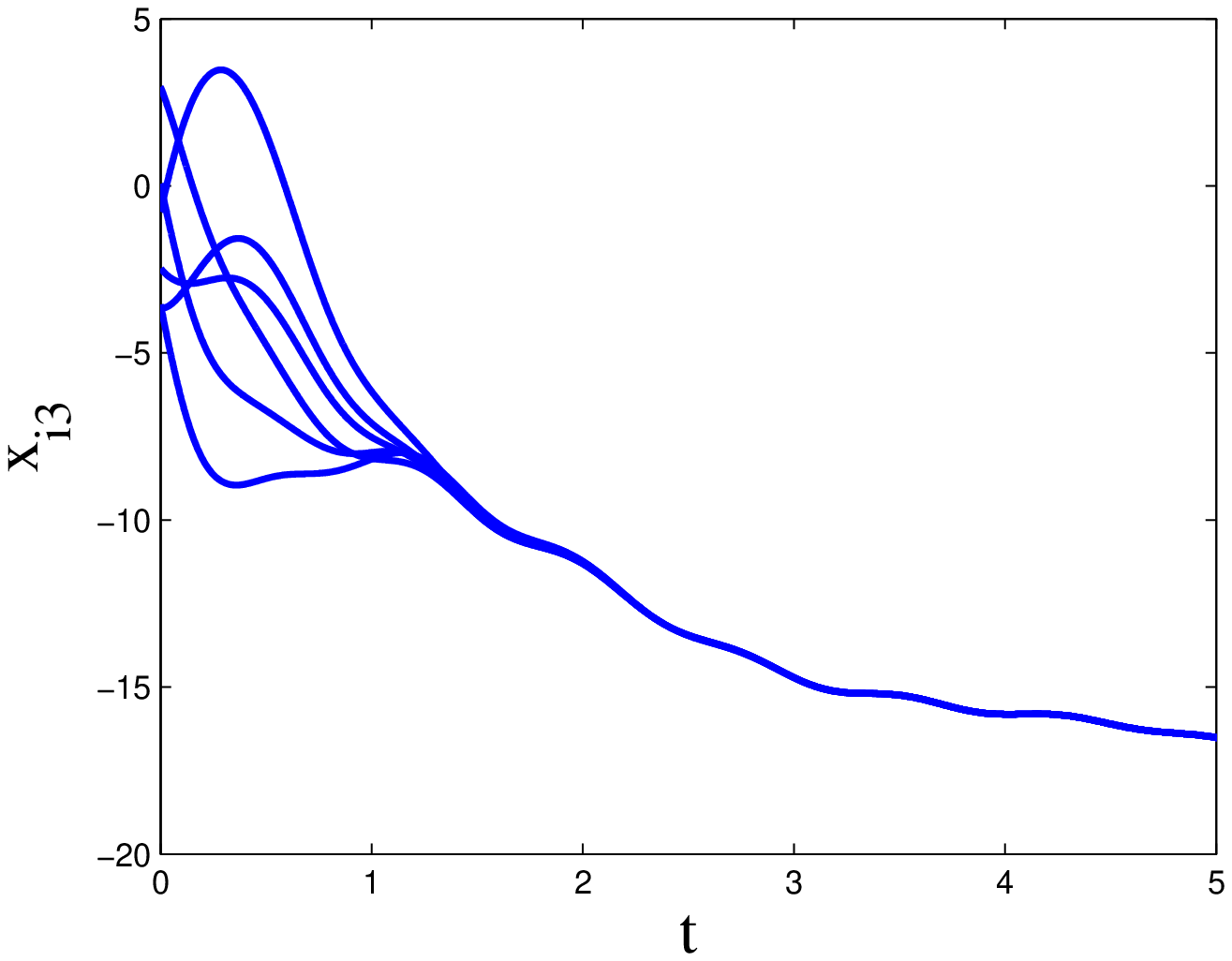}
\end{minipage}%
\begin{minipage}[b]{0.4\linewidth} \centering
\includegraphics[width=\linewidth,height=0.5\linewidth]{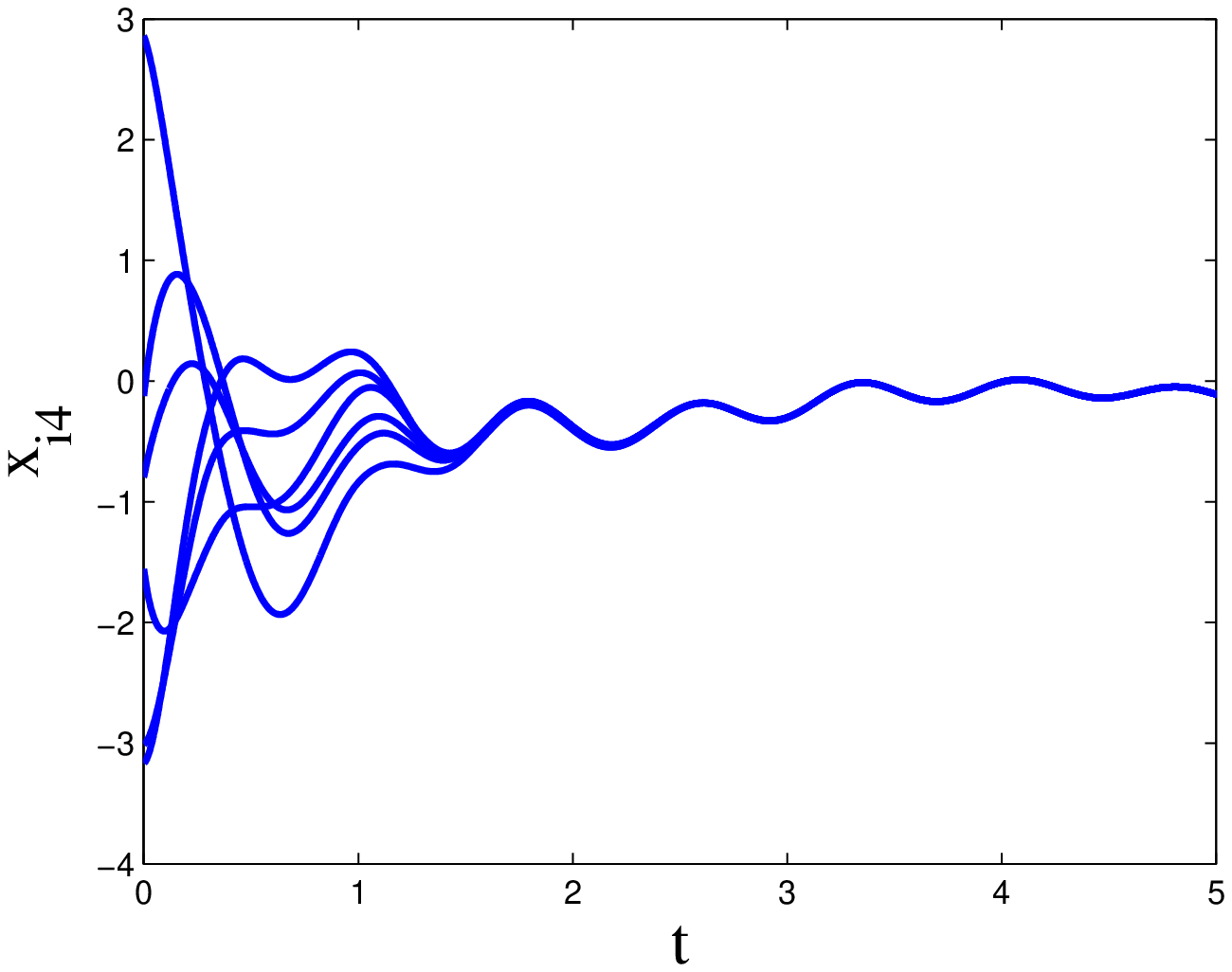}
\end{minipage}
\caption{The six manipulators reach global consensus.}
\end{figure}

Choose the $H_\infty$ performance index $\gamma=2$.  Solving the LMI
\dref{t51} by using the LMI toolbox of Matlab gives a feasible
solution:
$$\begin{aligned}
P=\begin{bmatrix}
0.4060 &  -0.9667  &  0.3547 &  -0.0842\\
-0.9667 &  67.6536 &   0.0162 &  -0.0024\\
0.3547  &  0.0162   & 0.4941  & -0.0496\\
-0.0842 &  -0.0024  & -0.0496  &  0.0367
\end{bmatrix},\quad \epsilon=29.6636.
\end{aligned}$$
Thus, by Algorithm 2, the feedback gain matrix of \dref{cl} is
chosen as
$$K=\begin{bmatrix} -2.4920 &  -0.1957  &  1.4115 &  -3.8216\end{bmatrix}.$$
For illustration, let the communication graph $\mathcal {G}$ be
given as in Fig. 1. It is easy to verify that $\mathcal {G}$
is balanced and strongly connected.
The corresponding Laplacian matrix is
$$
\mathcal {L}=\begin{bmatrix} 2 & 0 & -1& -1 & 0 &0\\
 -1 & 2 & 0& 0& 0& -1\\ 0 &-1 &1 &0 &0 &0\\ -1 & 0 & 0 & 2 & -1 &0\\
 0 & 0&  0& -1 &2 &-1\\ 0 &-1 &0 &0 &-1 &2\end{bmatrix}.
$$
The smallest nonzero eigenvalue of $\frac{\mathcal {L}+\mathcal {L}^T}{2}$
is equal to 0.8139. By
Theorem 2 and Algorithm 2, the protocol \dref{cl} with $K$ chosen
as above achieves global consensus with a $H_\infty$ performance
$\gamma=2$, if the coupling strength $c\geq 36.4462$. For the case
without disturbances, the state trajectories of the six manipulators
under the protocol \dref{cl} with $K$ given as above and $c=37$ are
depicted in Fig. 2, from which it can be observed that the global
consensus is indeed achieved. With the zero-initial condition and
external disturbances $\omega$, the trajectories of the performance
variables $z_i$, $i=1,\cdots,6$, are shown in Fig. 3.

\begin{figure}[htbp]
\centering
\includegraphics[width=0.35\linewidth]{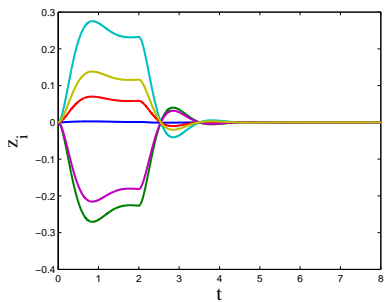}
\vspace*{-10pt} \caption{The performance variables $z_i$,
$i=1,\cdots,6$. }
\end{figure}

\section{Conclusion}

This paper has considered the global consensus problems of a class
of nonlinear multi-agent systems with Lipschitz nonlinearity
and directed communication graphs. A
two-step algorithm has been presented to construct a relative-state
consensus protocol, under which a Lipschitz multi-agent system without
disturbances can reach global consensus for a strongly connected directed
communication graph. Another algorithm has been then given to design
a protocol which can achieve global consensus with a guaranteed
$H_\infty$ performance for a Lipschitz multi-agent system subject to
external disturbances. 
An interesting
topic for future research is to investigate the cooperative control
problems of other types of nonlinear
multi-agent systems.

%

%
%

\end{document}